\documentclass[journal]{IEEEtran}
\usepackage{cite}

\usepackage{color}

\usepackage{array}

\usepackage[switch]{lineno}

\ifCLASSINFOpdf
   \usepackage[pdftex]{graphicx}
   \usepackage{subfigure}
   \graphicspath{{../pdf/}{../jpeg/}}
   \DeclareGraphicsExtensions{.pdf,.jpeg,.png}
\else
   \usepackage[dvips]{graphicx}
   \usepackage{subfigure}
   \graphicspath{{../eps/}}
   \DeclareGraphicsExtensions{.eps}
\fi
\begin{document}
%
\title{Response of the BGO Calorimeter to Cosmic Ray Nuclei in the DAMPE Experiment on Orbit}
%
%
%

\author{H.T. Dai, Y.L. Zhang*, J.J. Zang, Z.Y. Zhang, Y.F. Wei, L.B. Wu, C.M. Liu, C.N. Luo, D. Kyratzis, A. De Benedittis, C. Zhao, Y. Wang, P.C. Jiang, Y.Z. Wang, Y.Z. Zhao, X.L. Wang, Z.Z. Xu, G.S. Huang* 
\thanks{H.T. Dai, Y.L. Zhang, Z.Y. Zhang, Y.F. Wei, L.B. Wu, C.M. Liu, C. Zhao, Y. Wang, P.C. Jiang, Y.Z. Wang, Y.Z. Zhao, X.L. Wang, Z.Z. Xu and G.S. Huang are with the State Key Laboratory of Particle Detection and Electronics,  Department of Modern Physics, University of Science and Technology of China, Hefei, Anhui 230026, China. J.J. Zang and C.N. Luo are with the Key Laboratory of Dark Matter and Space Astronomy, Purple Mountain Observatory, Chinese Academy of Sciences, Nanjing, 210033, China. D. Kyratzis is with Gran Sasso Science Institute (GSSI), Via F. Crispi 7, I-67100 L'Aquila, Italy and INFN Istituto Nazionale di Fisica Nucleare, Lab. Nazionali del Gran Sasso, Via Acitelli, I-67100 Assergi (L'Aquila), Italy. A. De Benedittis is with Dipartimento di Matematica e Fisica E. De Giorgi, Universita\`\ del Salento, I-73100, Lecce, Italy and Istituto Nazionale di Fisica Nucleare (INFN) - Sezione di Lecce, I-73100, Lecce, Italy. (Corresponding author e-mail: ylzhang@ustc.edu.cn, hgs@ustc.edu.cn)}
}

%
%

\markboth{O30-1-5}%
{Shell \MakeLowercase{\textit{et al.}}: Bare Demo of IEEEtran.cls for IEEE Journals}
%



\maketitle

\begin{abstract}
DArk Matter Particle Explorer (DAMPE), the satellite-based cosmic ray and gamma ray measurement experiment, relies on its calorimeter to measure the energy of incident particles. The calorimeter adopts crystals of bismuth germanium oxide (BGO) as scintillating material, and it is designed to aim for measurements of energy ranging from 50 GeV to 100 TeV in the case of a cosmic ray nucleus. This work concerns the response of the BGO calorimeter to nucleus-type cosmic rays. Cosmic rays with very low energy can rarely reach the detector due to the Earth's magnetic field. A cutoff on lower energy can be observed in the energy spectrum. In this work, the cutoff is used to study the response of the calorimeter. Carbon, neon, silicon and iron are analyzed separately in comparison with Monte Carlo simulations by Geant4.
\end{abstract}

\begin{IEEEkeywords}
BGO calorimeter, CR Nuclei, geomagnetic cutoff, DAMPE, energy response
\end{IEEEkeywords}

%
\IEEEpeerreviewmaketitle

\section{Introduction}
%
%
%
%
\IEEEPARstart{F}{or} space experiments such as Fermi Large Area Telescope (LAT) \cite{FermiLAT}, CALorimetric Electron Telescope (CALET) \cite{CALET} and DArk Matter Particle Explorer (DAMPE) \cite{chang2017dark}, the detectors need to be calibrated with well-known sources of astrophysical origin. Meanwhile, even with a complete calibration there are still too many factors that might influence the measurement and the reconstruction of science data. It is meaningful to make sure that our knowledge about the response of our instrument is reliably reflecting the reality. In order to achieve this, the geomagnetic cutoff feature in cosmic ray (CR) spectrum can be used as a source to investigate the performance of the calorimeter. A brief introduction is given as follows.

On the orbit of DAMPE (sun-synchronous orbit with inclination of ~97 degrees and altitude of 500 km), cosmic rays with low momentum can hardly be observed, because they are bent by the Earth's magnetic field when radiating towards the Earth. The trajectory of a charged particle in a magnetic field follows the equation:

\begin{equation}
R = P/z \approx 0.3 \times B \times \rho
\end{equation}

where R is the rigidity, P is the momentum of the particle in GeV/c, z is the charge number (absolute value), B is magnetic flux density in Tesla and $\rho$ is the radius in meters. This sets R in the unit of GV, representing momentum per unit of charge. The equation tells us that cosmic rays carrying the same charge are more easily bent with lower momentum. At different position over the Earth, the geomagnetic cutoff in the spectrum, viewed by the DAMPE spectrometer, varies largely from $\sim$1 GV to $\sim$12 of GV. This makes the cutoff on the energy of cosmic ray nuclei extend from several GeV to hundreds of GeV. For example, CR iron acquired near the equator has the cutoff on energy over 200 GeV.

This cutoff in the energy spectrum can also be determined numerically by tracing the CR nuclei in the magnetic field, and a comparison between the measured data and the simulated ones can be performed.

Similar work on cosmic ray electron-plus-positron (CRE) has been done by Fermi-LAT \cite{EnergyScaleLAT}. The major differences from this previous work are focused on two specific points: firstly the cutoff in CR nuclei spectrum is higher because particles that carry more charge are more strongly shielded by the Earth's magnetic field, i.e. nuclei with higher atomic number have higher cutoff on their energy. Secondly in the CR's interaction with the atmosphere rarely are heavy ions produced, which means the cutoff in the spectrum has no secondary contamination.

\begin{figure}[!t]
\centering
\includegraphics[width=3.5in]{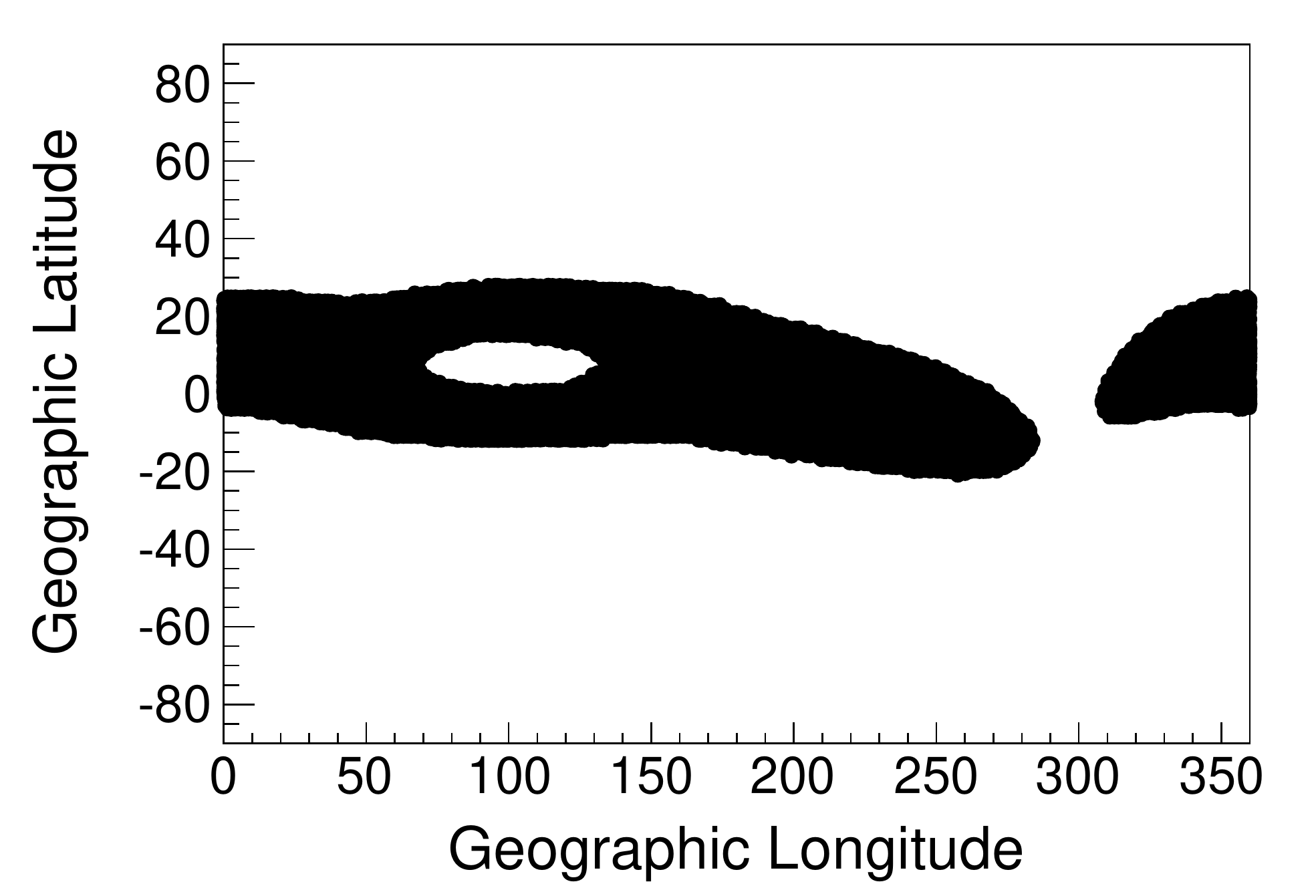}
\caption{The region of McIlwain L value between 1.00 and 1.14 for the orbit at an altitude of 500 km.}
\label{mapL}
\end{figure}

The cutoff value varies at different positions near the Earth. Roughly speaking, it reaches its maximum near the equator. The analysis in this paper is performed using data collected within the shaded area in Fig. \ref{mapL} to study the response of the BGO calorimeter in higher energy. It is where we can get a maximum cutoff and enough data in the mean time. This area is defined by the McIlwain L interval 1.00-1.14. The McIlwain L value is a parameter describing the density of magnetic field lines crossing the Earth's magnetic equator, thus being an appropriate way to characterize the geomagnetic cutoff \cite{Smart2005}.

\begin{figure}[h]
\centering
\includegraphics[width=3.5in]{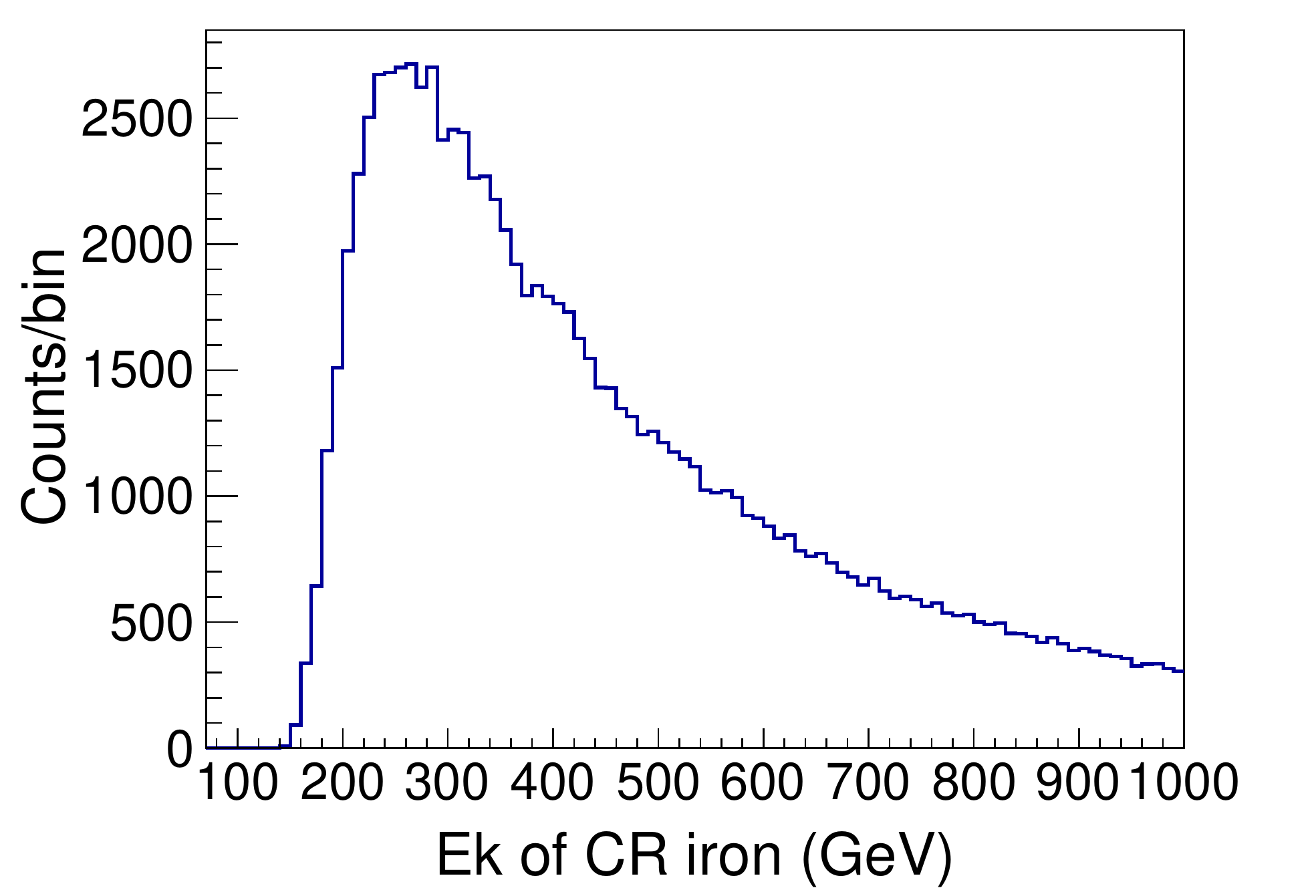}
\caption{The distribution of kinetic energy of CR irons that can cross the DAMPE orbit in the area of McIlwain L interval (1.00, 1.14), uniform binning adopted. It is a result of "toy" simulation, obtained simply by simulating the geomagnetic field effect on CR irons without considering any detector response. From this we can see a clear structure of cutoff by geomagnetic effect and it is located at over 200 GeV for CR iron.}
\label{cutoff_Fe}
\end{figure}

Due to the geomagnetic field, figure \ref{cutoff_Fe} represents a typical distribution of the kinetic energy of cosmic rays. The spectrum, obtained by only simulating the geomagnetic effect, shows the energy of CR iron that can cross the DAMPE orbit within this area. The counts of CR iron rise from \textasciitilde150 GeV and reach maximum at \textasciitilde250 GeV, then slowly drop as energy goes higher. Without atmospheric contamination at low energy, the cutoff is more clearly shown and at much higher energy than the cutoff on CRE presented in \cite{EnergyScaleLAT}. Thus it can be used as a good reference to study the response of the calorimeter to particles with higher energy. In the spectrum, the left shoulder is a rising edge instead of a cutoff at a single value, because cosmic rays are not observed at one single position over the Earth, and they can reach the detector from different directions.

\section{DAMPE Spectrometer}
\subsection{The structure of DAMPE}
\begin{figure}[h]
\centering
\includegraphics[width=3.5in]{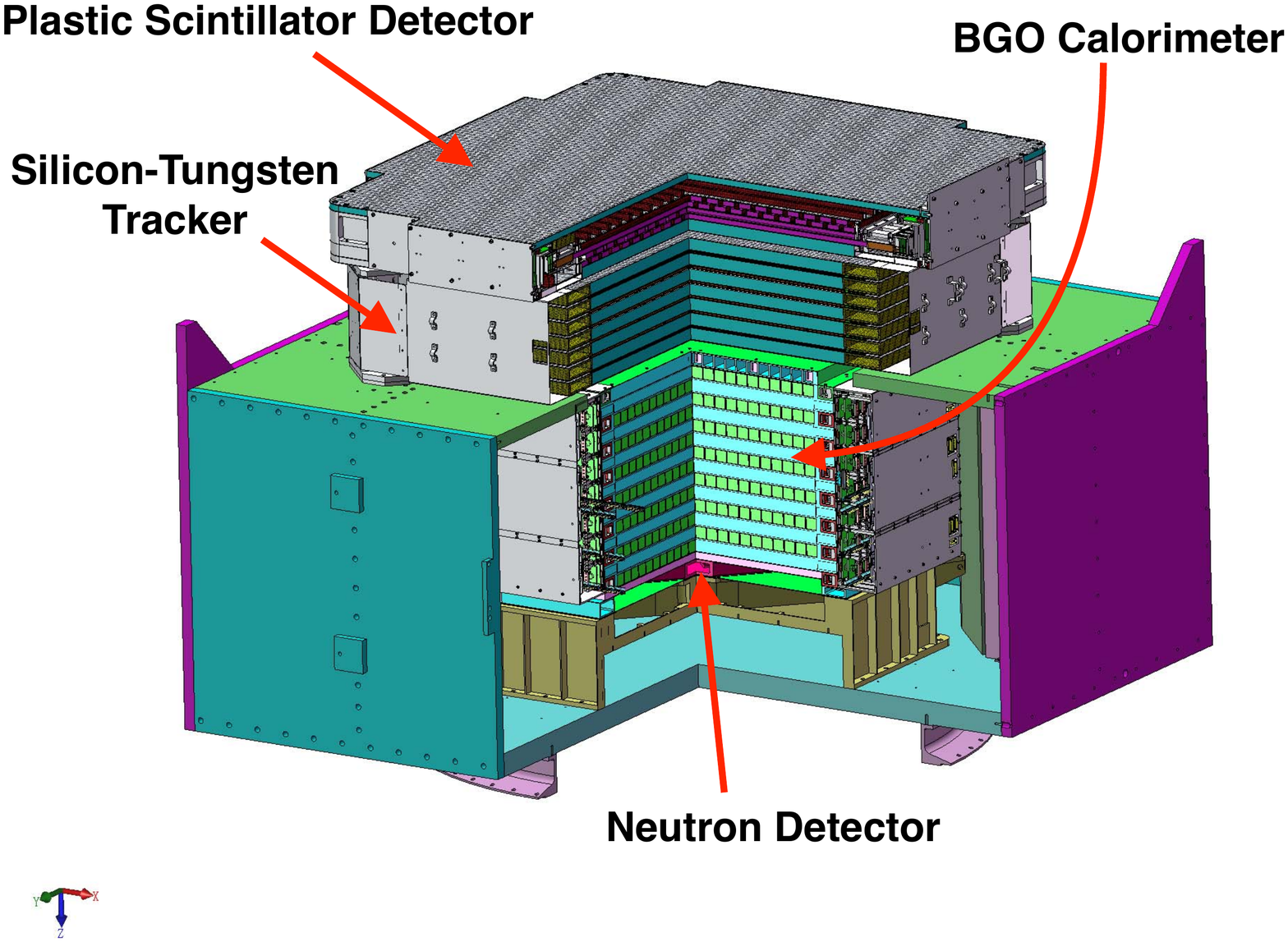}
\caption{A schema of DAMPE.}
\label{DAMPE structure}
\end{figure}

DAMPE is a satellite-based telescope aiming at detection of very high energy cosmic rays and gamma rays. Figure \ref{DAMPE structure} is a schema of it. The whole detector consists of a Plastic Scintillator Detector (PSD), a Silicon-Tungsten tracKer-converter (STK), a BGO imaging calorimeter (BGO), and a NeUtron Detector (NUD) from top to bottom \cite{chang2017dark}. The PSD is essentially utilized in order to provide the charge number (for CR nuclei it's also the atomic number $|Z|$) of incident particles, as well as being an anti-coincidence detector for $\gamma$-rays. The STK reconstructs the trajectory. The BGO calorimeter measures the energy and distinguishes electromagnetic particles from hadrons. The BGO image also gives some rough track information. The NUD provides additional electron-hadron discrimination, which is important for energy range above TeV. The four subdetectors above provide good measurements of the charge, arrival direction, energy, and particle identification to accomplish major science objectives of DAMPE, including indirect search for dark matter signals, $\gamma$-ray astronomy, and studies on the origin and propagation mechanism of Galactic Cosmic Rays (GCRs).

\subsection{The BGO calorimeter and energy measurement}

\begin{figure}[h]
\centering
\includegraphics[width=3.5in]{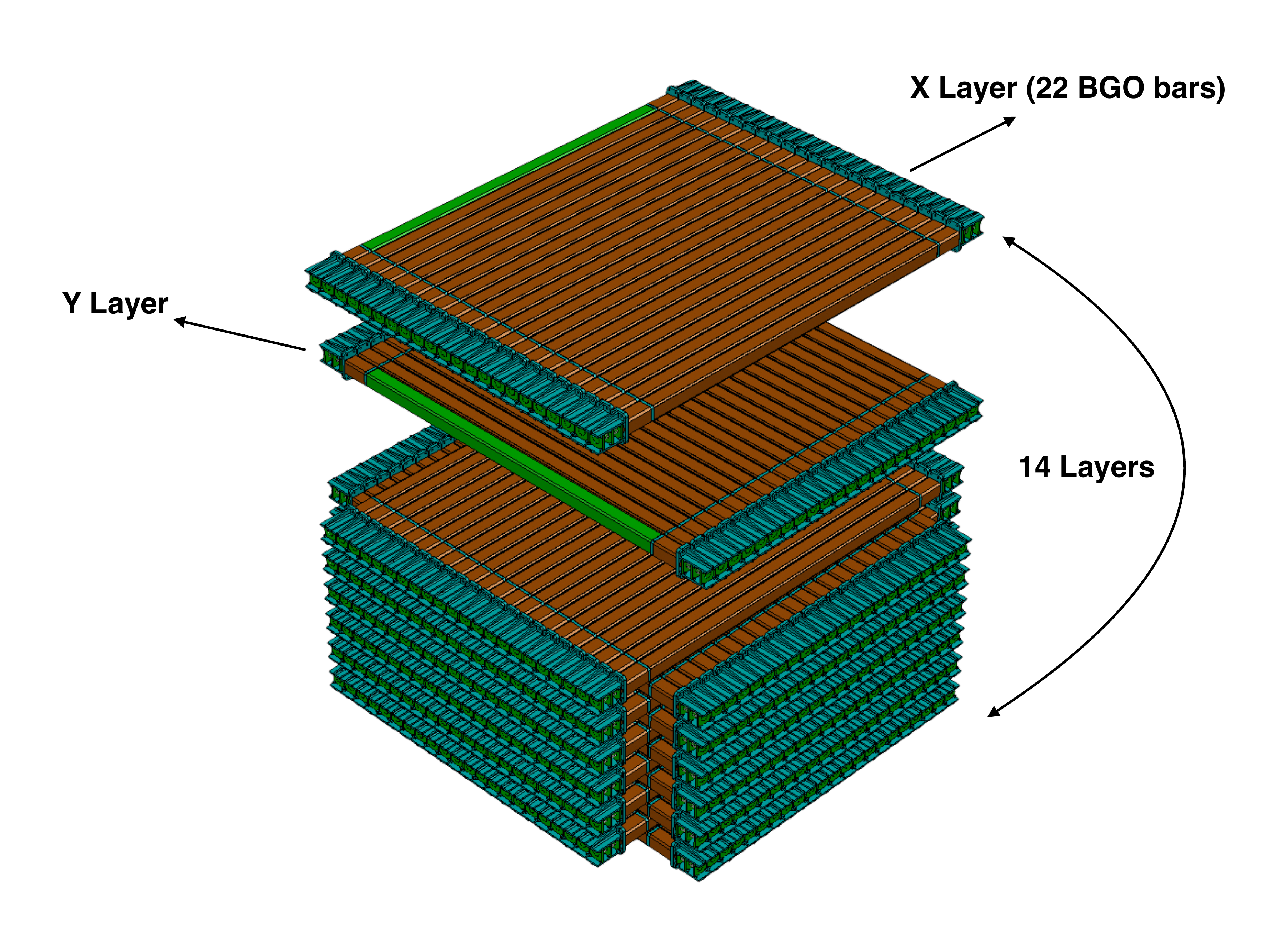}
\caption{The calorimeter consists of 14 layers. Each layer has 22 BGO bars in x or y direction.}
\label{BGO_structure}
\end{figure}

The calorimeter contains 14 layers, each of 22 BGO crystal bars that are arranged alternately in x or y direction within each layer, as shown in figure \ref{BGO_structure}. All of the 308 BGO bars are of size 25 mm $\times$ 25 mm $\times$ 600 mm. The calorimeter is of 1.6 $\lambda_I$ (nuclear interaction length) from top to bottom, which is crucial to the energy measurement of CR nuclei. 

To validate the fidelity of the instrument model and the simulation, beam tests were performed in 2014 and 2015 at CERN, with high energy gamma rays, electrons, protons, muons and various nuclei produced by fragmentation of Argon and Lead on the Engineering Qualification Model (EQM) of DAMPE \cite{chang2017dark}.

The on-orbit calibration for energy measurement consists of the pedestal calibration, the zero-suppression threshold and electronics linearity, the MIP response calibration, the PMT dynode ratio calibration, and the light attenuation calibration \cite{ambrosi2019orbit}. The MIP signals by relativistic protons are compared with the distribution of deposited energies given by Monte Carlo simulations of the on-orbit spectrum of CR protons that should be detected by DAMPE. This gives the parameter of the transfer function that converts signals in digital counts to energies that a particle releases in the crystal \cite{wu2018calibration}.

\section{Simulation of the Geomagnetic Cutoff}
Particles that can reach DAMPE are subject to the primary spectrum of some certain type of CR modified by the Earth's magnetic field. When doing the simulation, the primary spectrum measured by other previous experiment is used as an input to Geant4, thus the DAMPE response simulated. The effect of the geomagnetic field is considered in the next step. The two steps  are separately introduced as follows.

\subsection{Primary spectrum of CR nuclei and Geant4 simulation}
As mentioned in the introduction, the cutoff on energy increases as the charge number of CR nuclei gets larger. Carbon, neon, silicon and iron are four elements chosen for investigating the BGO calorimeter response to the spectrum of geomagnetic cutoff. They are relatively abundant in cosmic rays, and the cutoff in their energy spectrum is distinct enough from each other, so they can be used as four different sources.

The experimental result of PAMELA was adopted as the primary spectrum of CR carbon to be put into Geant4 simulation \cite{carbonPAMELA}. Simulation for neon used the results of HEAO-3 C2 experiment \cite{neonHEAO3C2}, silicon and iron of ATIC02 \cite{Panov2009}. In this step, the response of the whole DAMPE detector to the primary CR nuclei were given via Monte Carlo simulations. Then the sample were to be further simulated for a real orbit detection, using the so-called back tracing technique, within the region of McIlwain L interval 1.00-1.14. Thus the effect of geomagnetic field would be considered.

\subsection{Back tracing in the magnetic field}
The International Geomagnetic Reference Field models (IGRF) serve as a series of standard descriptions of the Earth's magnetic field \cite{IGRF}. Mathematically these models are spherical harmonic expansions of the geomagnetic potential. The version of IGRF-12 was adopted for this work. Generally, other external sources (such as solar wind) contribute little to shielding the Earth compared to the Earth's internal magnetic field \cite{zuccon2002monte}, though the real magnetic environment around the Earth is a multisource system \cite{ModelingMagnetopause}, especially when one moves farther away from the Earth. Data collected during strong geomagnetic storms \cite{MagnetosphereDuringStorms} are excluded in the analyses, which further reduces the influence from dynamic external sources of the field.

As it is hard to trace a particle coming from the Galaxy in the Earth's magnetic field and expect it to collide with the DAMPE detector, back tracing a particle from the coordinates of DAMPE to see whether it comes from the Galaxy or not would be a more practical way. A particle can be back traced from the position where the DAMPE satellite is located, and if it turns out that it intersects the Earth or is captured by the Earth (back traced for a given time that is long enough and pointless to compute more), it is considered unphysical, as it cannot be a primary GCR anymore. The particle is considered Galactic when it reaches 10 Earth radii. The code developed by Smart and Shea was used to compute the particle trajectory tracing \cite{Smart2005}. More detailed description of this method can be found in \cite{EnergyScaleLAT}.

We simulated the DAMPE orbit from Jan. 2016 through July 2017, and distributed some simulated events obtained by Geant4 on the location of every one second on this orbit, then back traced these events. Only the orbit in the region of McIlwain L interval 1.00-1.14 was considered.

\section{Performance of the Calorimeter}
\subsection{Selection of data}
\begin{figure}[h]
\centering
\includegraphics[width=8cm]{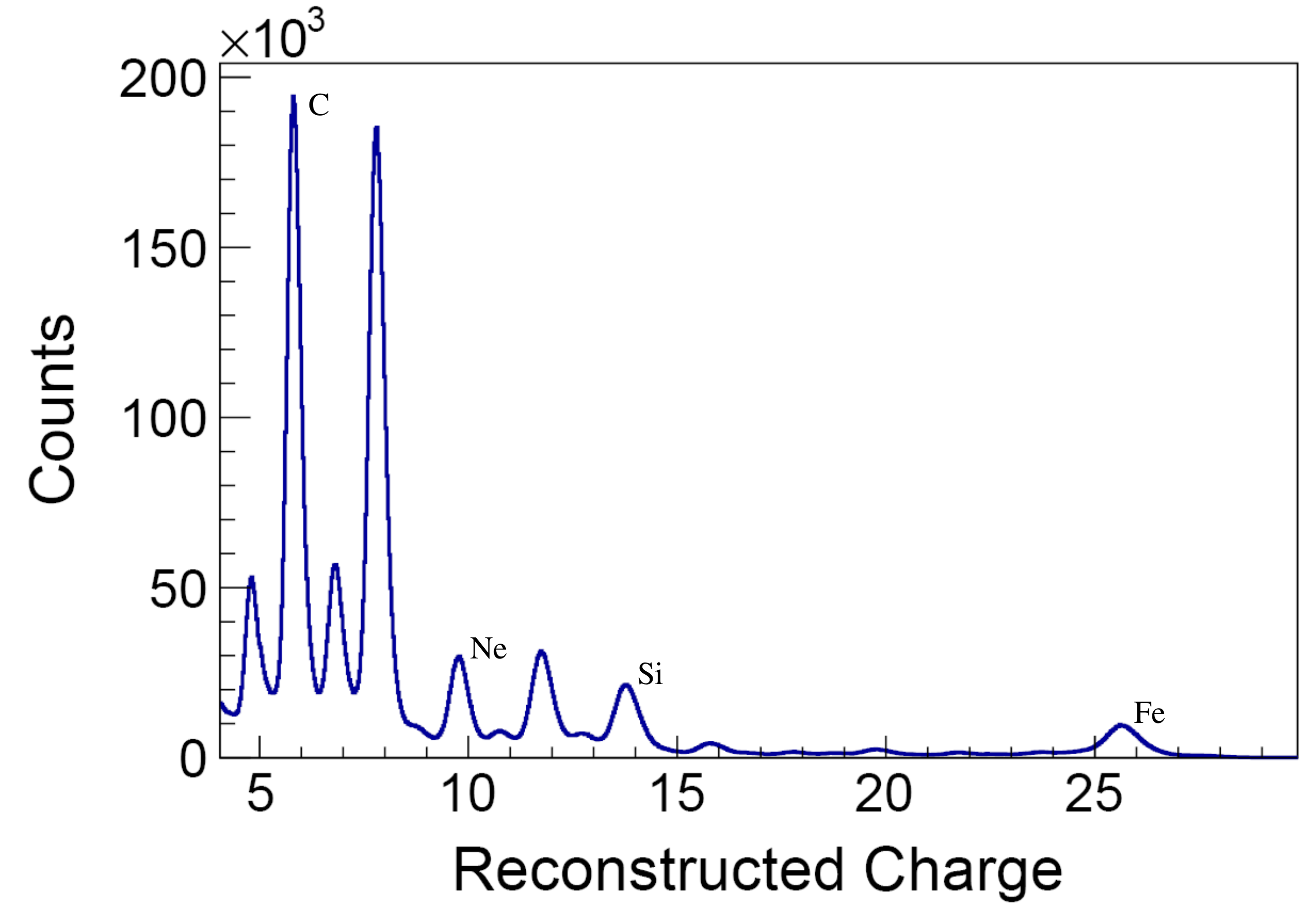}
\caption{The reconstructed charge by PSD subdetector. No correction is done here so that the center value of the peaks is not forced to be an integer. Carbon, neon, silicon and iron are picked out as four relatively abundant elements in CRs to do this analysis. Besides, they have the energy cutoff distant enough from each other.}
\label{ChargeAll}
\end{figure}

To distinguish different elements in CR, the event should at least have a reliable reconstructed charge in PSD subdetector. Figure \ref{ChargeAll} shows the charge spectrum of heavy ions detected by DAMPE. For each of the four peaks, the events within the full width at half maximum are taken as candidates for this element. 

To measure the energy of CR nuclei, it is better to select events with a sufficiently developed shower profile for the analysis. For this, we use only high-energy triggered events, and require the energy deposition maximum position to be in first 9 layers of BGO calorimeter, and not located too much on the side. Besides, the events are selected only when it deposits energy in the third layer 1.5 times larger than in the first layer. This condition makes sure that the cascade process sufficiently develops in the BGO calorimeter as soon as the particle enters. For the simulation data, all events that are classified as unphysical (that cannot be a real primary GCR) are eliminated for all the analyses in this paper.

\subsection{Validation of the energy deposition}
\begin{figure}[htbp]
\centering
\subfigure[energy in layer 2]{
\includegraphics[width=8cm]{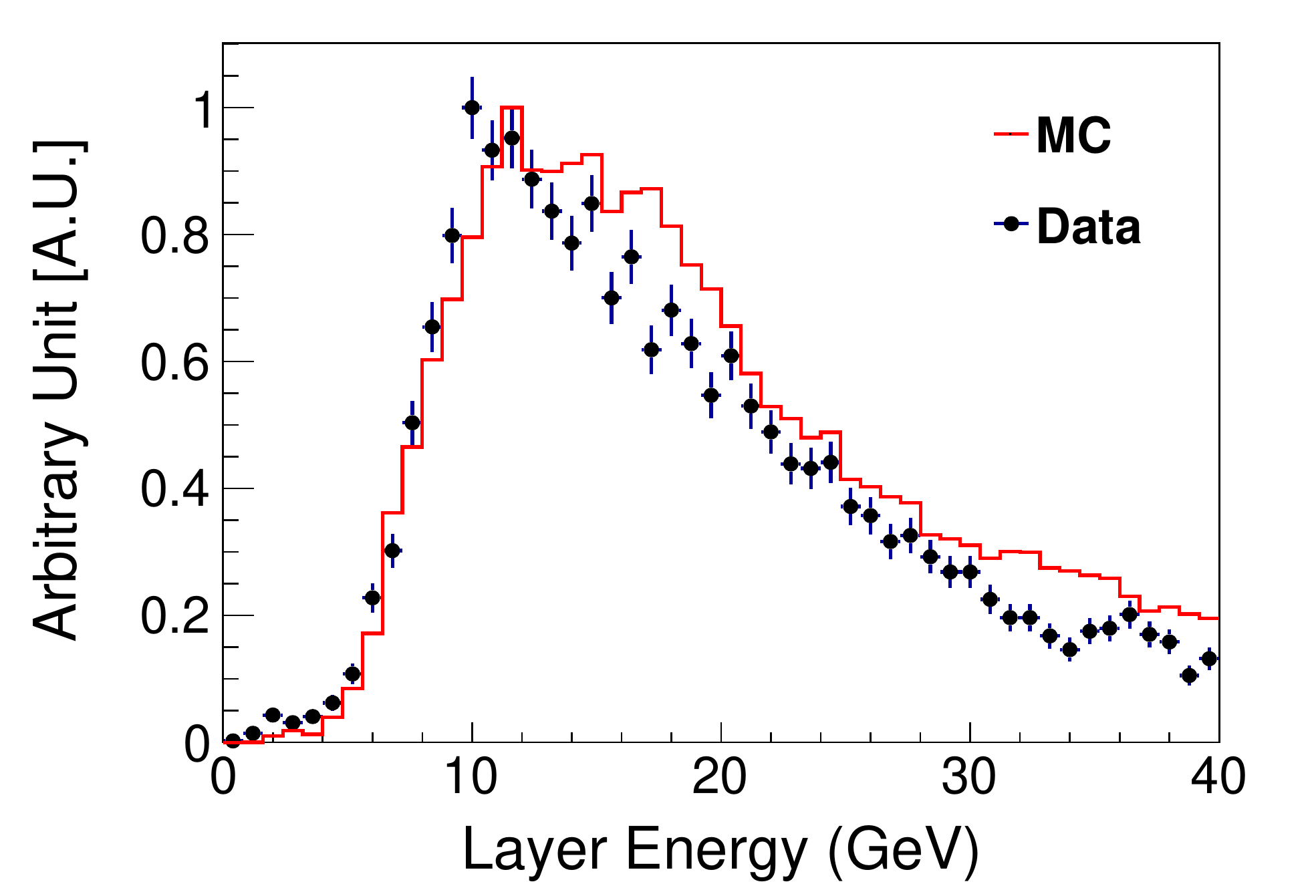}
}
\quad
\subfigure[energy in layer 5]{
\includegraphics[width=8cm]{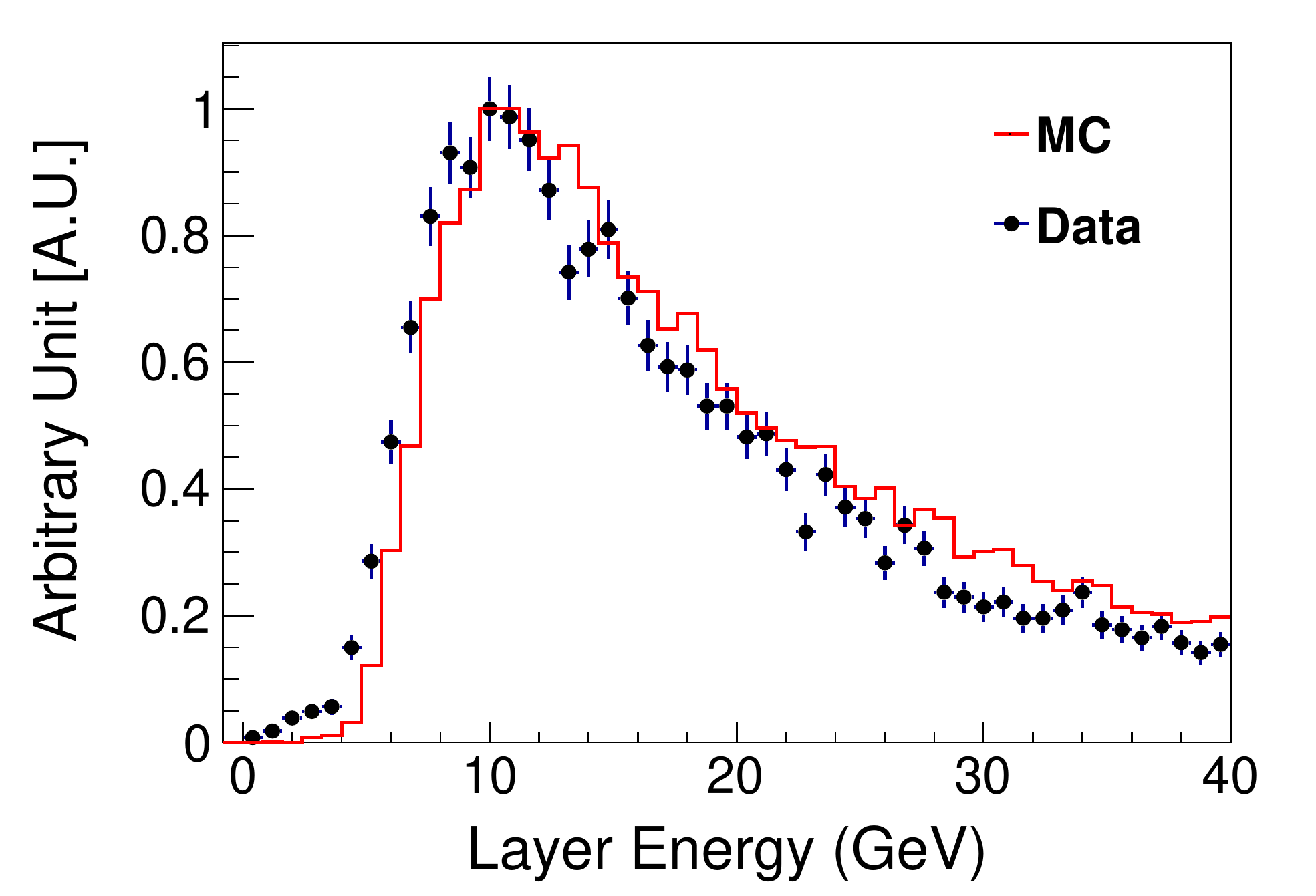}
}
\quad
\subfigure[energy in layer 8]{
\includegraphics[width=8cm]{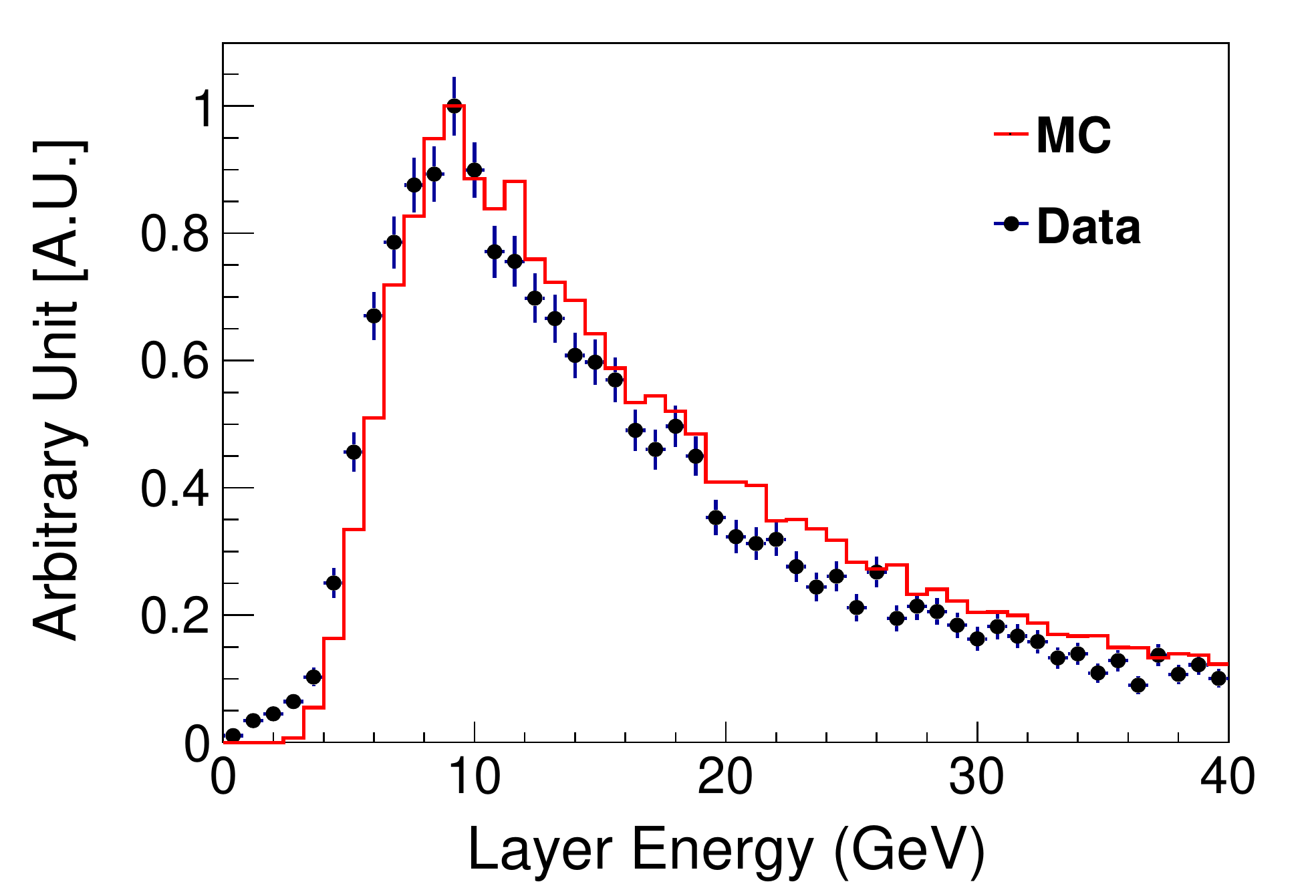}
}
\quad
\subfigure[energy in layer 11]{
\includegraphics[width=8cm]{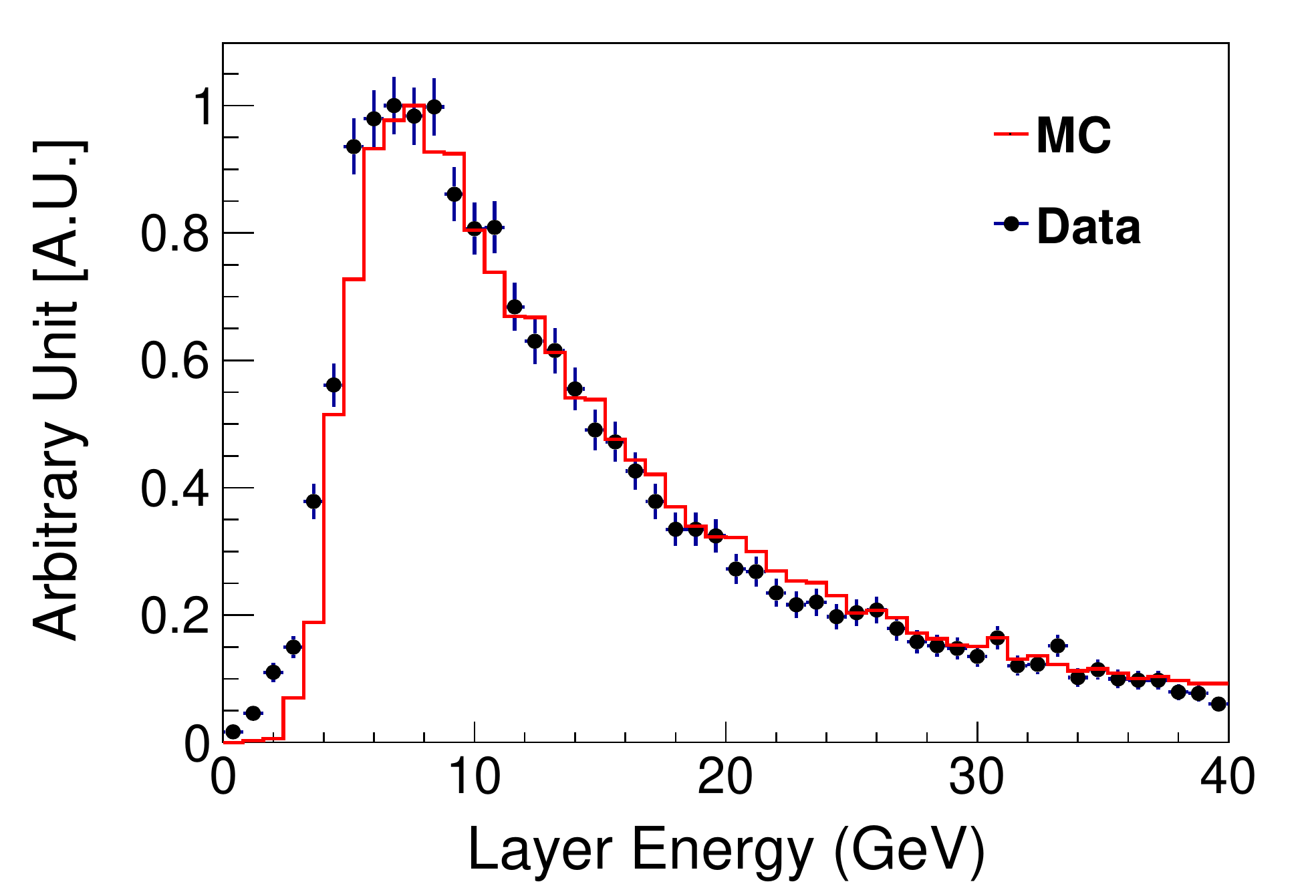}
}
\caption{The energy deposited in one layer of BGO calorimeter. Above are four layers among the total 14 to show the comparison between data and MC, respectively layer 2, 5, 8 and 11. Here we again take CR iron as an example.}
\label{LayerEnergy}
\end{figure}

It is the BGO calorimeter response to energy deposition that we want to study. Most of the effective events detected by DAMPE are coming from the top of BGO calorimeter because of the positioning of PSD and STK subdetectors, which provide information on the charge and track. Since the energy deposition in different layers from top to bottom reflects the longitudinal shower development, a layer-by-layer comparison between measured energy and simulation has been done. Figure \ref{LayerEnergy} is such a comparison on the energy deposition by CR iron, showing the situations within 4 single layers among the total 14 layers of the calorimeter. Approximately the profile of the distribution given by simulation agrees with the measured one, but still as the particle goes deeper in the calorimeter (as the shower sufficiently develops), they agree to a better level. This might come from the uncertainty from the interaction models when the shower hasn't sufficiently developed, where the secondary particles of cascade are relatively less than in the deep part of the calorimeter.
 As for in the layer 11 (the last case in figure \ref{LayerEnergy}), the fitted mean value of the distribution in flight data is $6.30\pm0.18$ GeV, while the fitted mean value is $6.52\pm0.16$ GeV for simulation. Within the errors, simulation agrees well with flight data.

\section{Response to CR Nuclei in the Energy Band around the Cutoff}
\subsection{Comparison of the total deposited energy}
\begin{figure}[htbp]
\centering
\subfigure[carbon]{
\includegraphics[width=8cm]{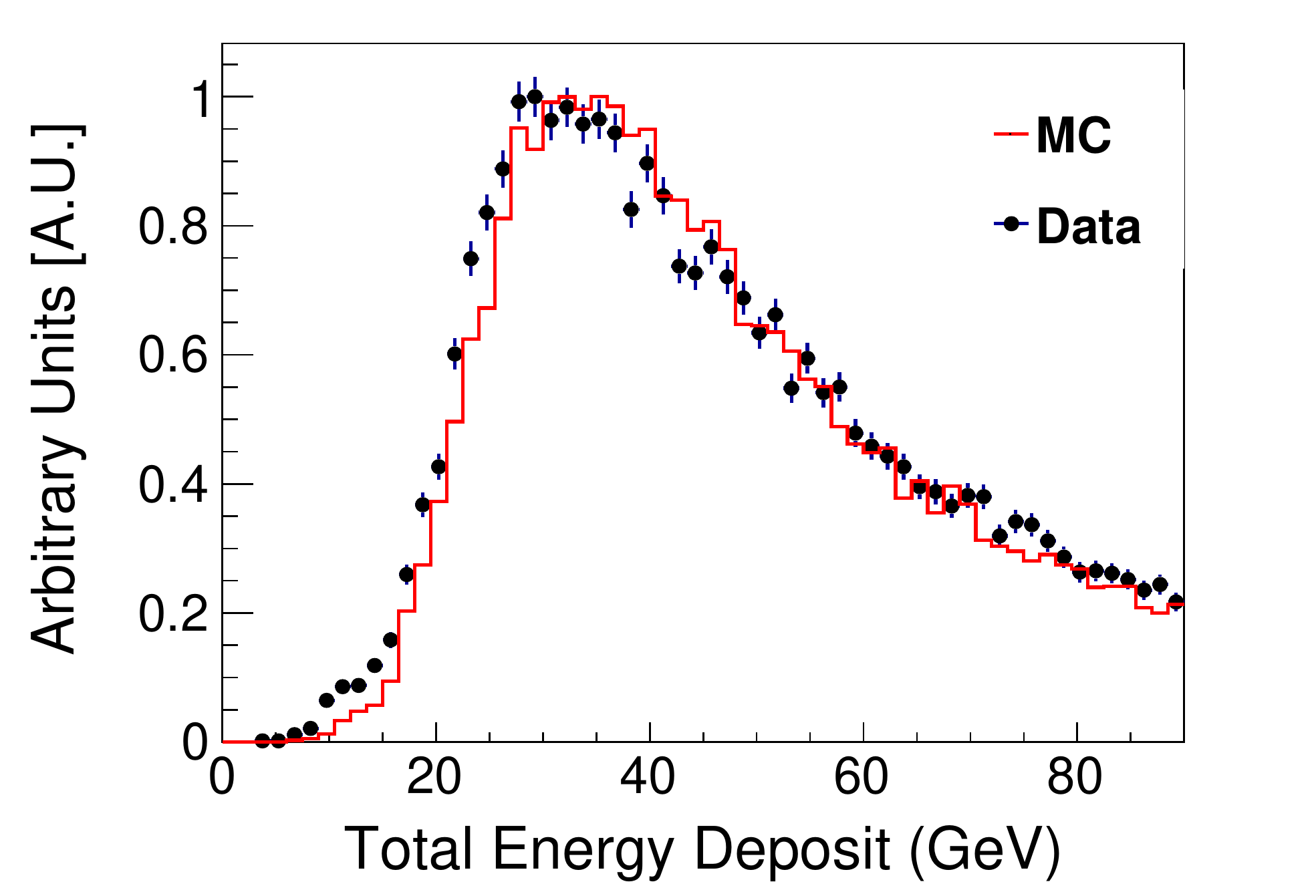}
}
\quad
\subfigure[neon]{
\includegraphics[width=8cm]{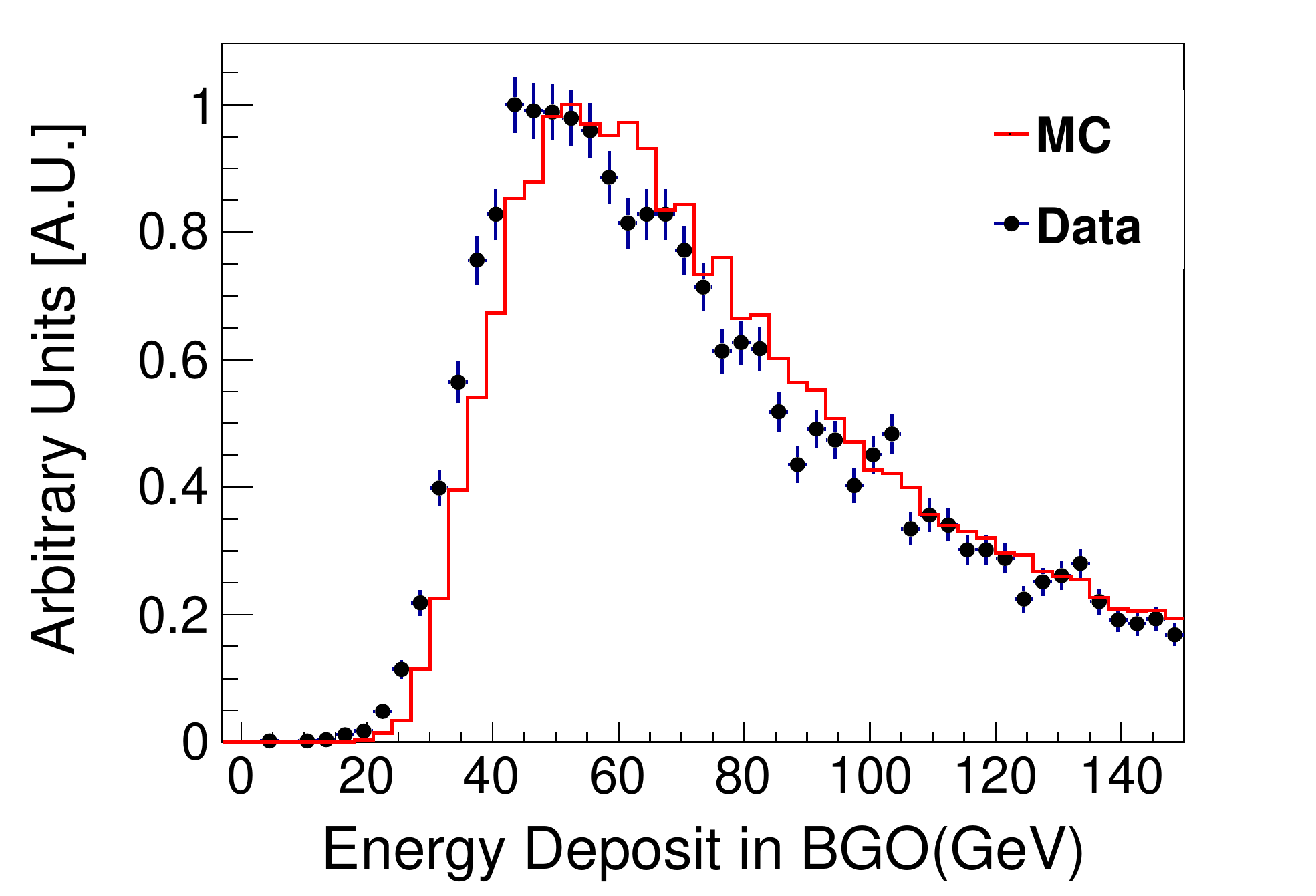}
}
\quad
\subfigure[silicon]{
\includegraphics[width=8cm]{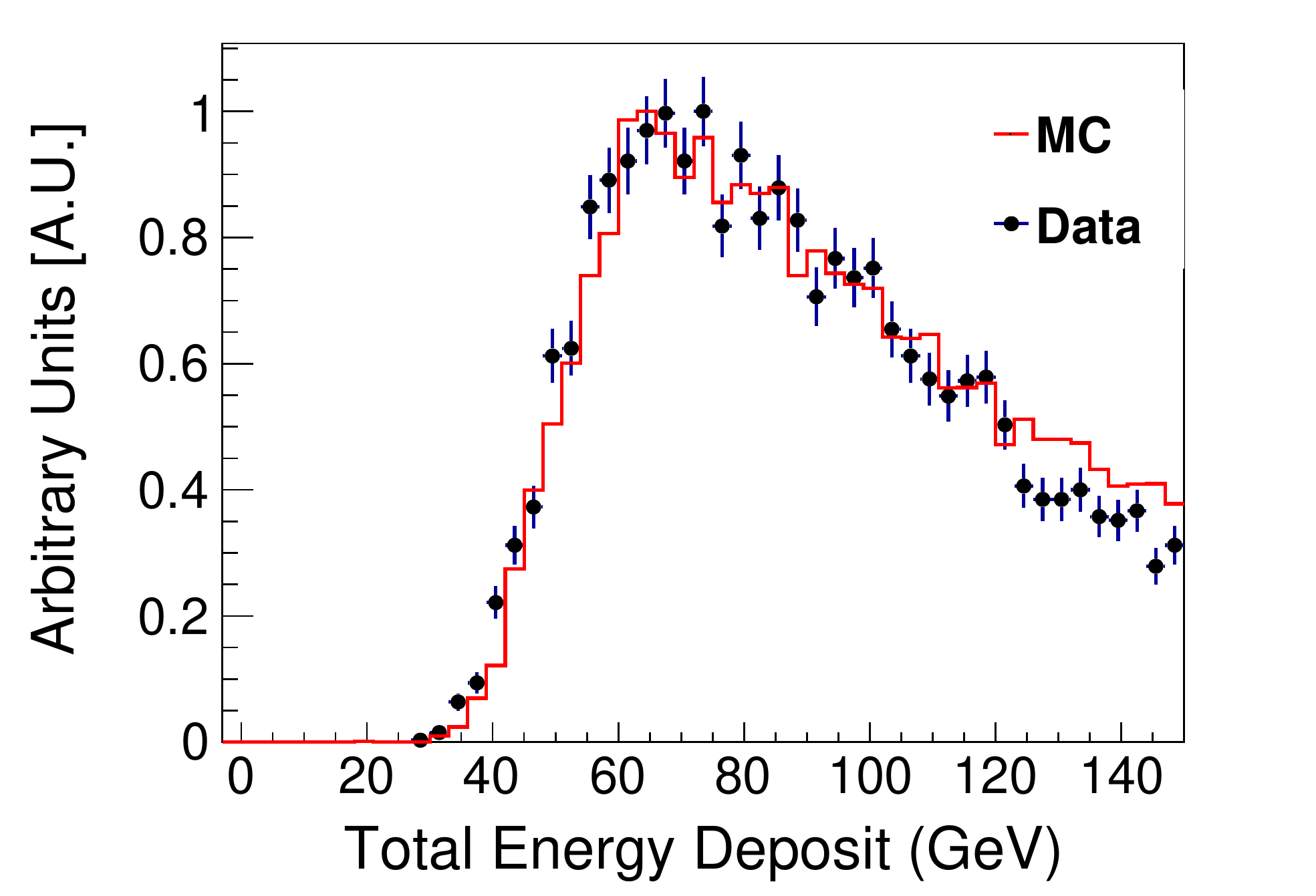}
}
\quad
\subfigure[iron]{
\includegraphics[width=8cm]{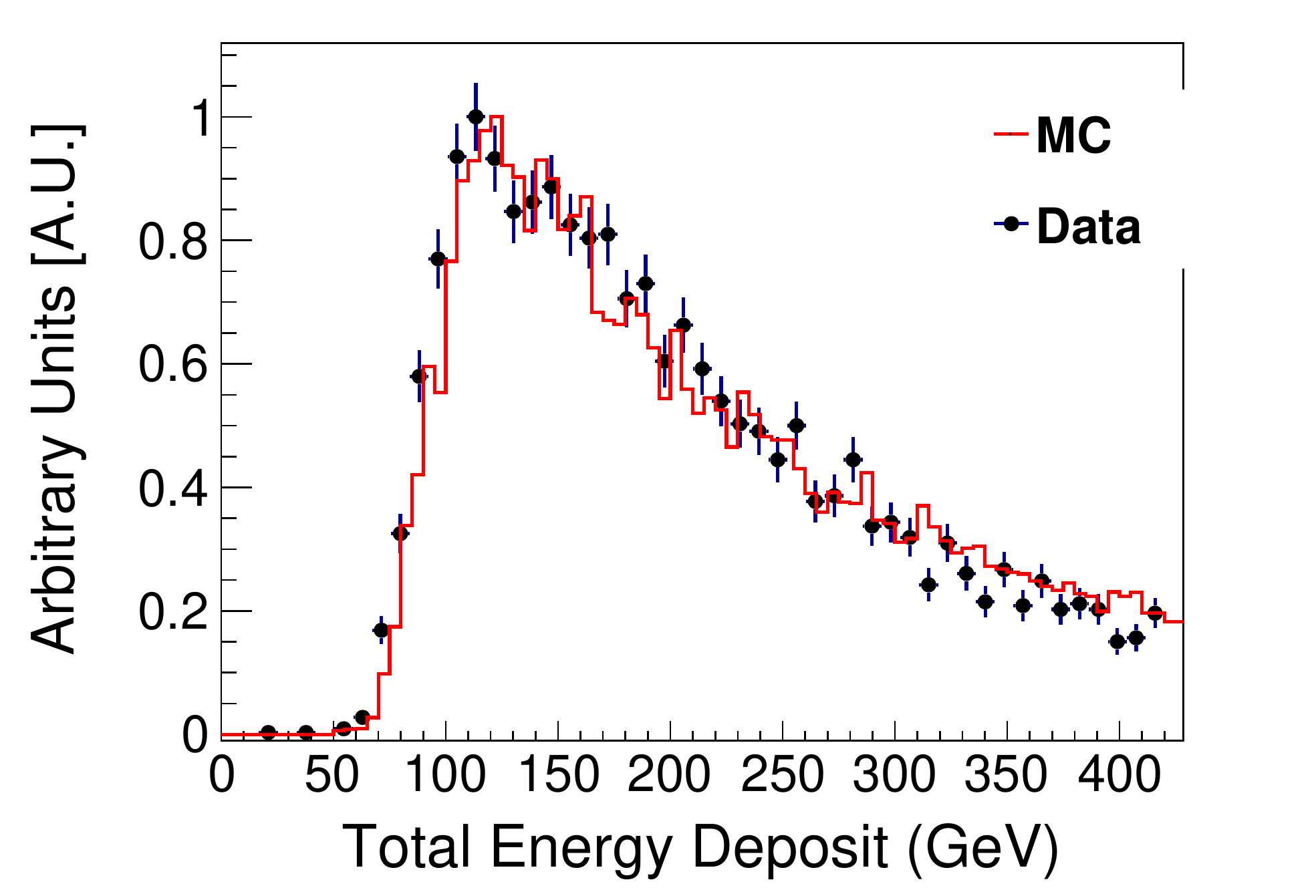}
}
\caption{ The energy deposited in the BGO calorimeter, respectively by CR carbon, neon, silicon and iron.  Flight data are set to dots and simulations are set to red lines to show the comparison between data and MC.}
\label{TotalEnergyComparison}
\end{figure}

The response of the whole BGO calorimeter to CR carbon, neon, silicon and iron can be investigated. According to the beam tests performed at CERN, the total deposited energy is estimated to be $\sim$30-40\% of the incident energy for nuclei \cite{chang2017dark}. The deposited energy spectrum of the selected data sample has a cutoff, which locates at roughly $\sim$30-40\% the physical energy cutoff of this CR nuclei. The spectra given by simulations and flight data are drawn together to give a visual comparison between them in Fig. \ref{TotalEnergyComparison}. Generally speaking, the profile of the distribution agrees well between simulations and flight data. The measured deposited energy fluctuates around the simulation, which means the response of the BGO calorimeter agrees with our simulation models to some extent. Still, the rising edge by flight data is located a little to the left of simulations, and the agreement of the rising edge gets better as heavier CR nuclei are analyzed.

\subsection{Fitting the count spectra}
The count spectra of energy in figure \ref{TotalEnergyComparison} can be parametrized by the function below,

\begin{equation}
dN(E)/dE = cE^{-\gamma}/(1+e^{-a(E-E_c)})
\end{equation}

where $\gamma$ is the spectral index and $E_c$ the cutoff on energy. $c$ stands for no special meaning but the magnitude of event counts. $a$ is a parameter representing the steepness of the rising edge in the spectrum. The spectra of deposited energy of flight data and simulations were fitted with the function separately.

\begin{figure}[htbp]
\centering
\subfigure[simulation]{
\includegraphics[width=8cm]{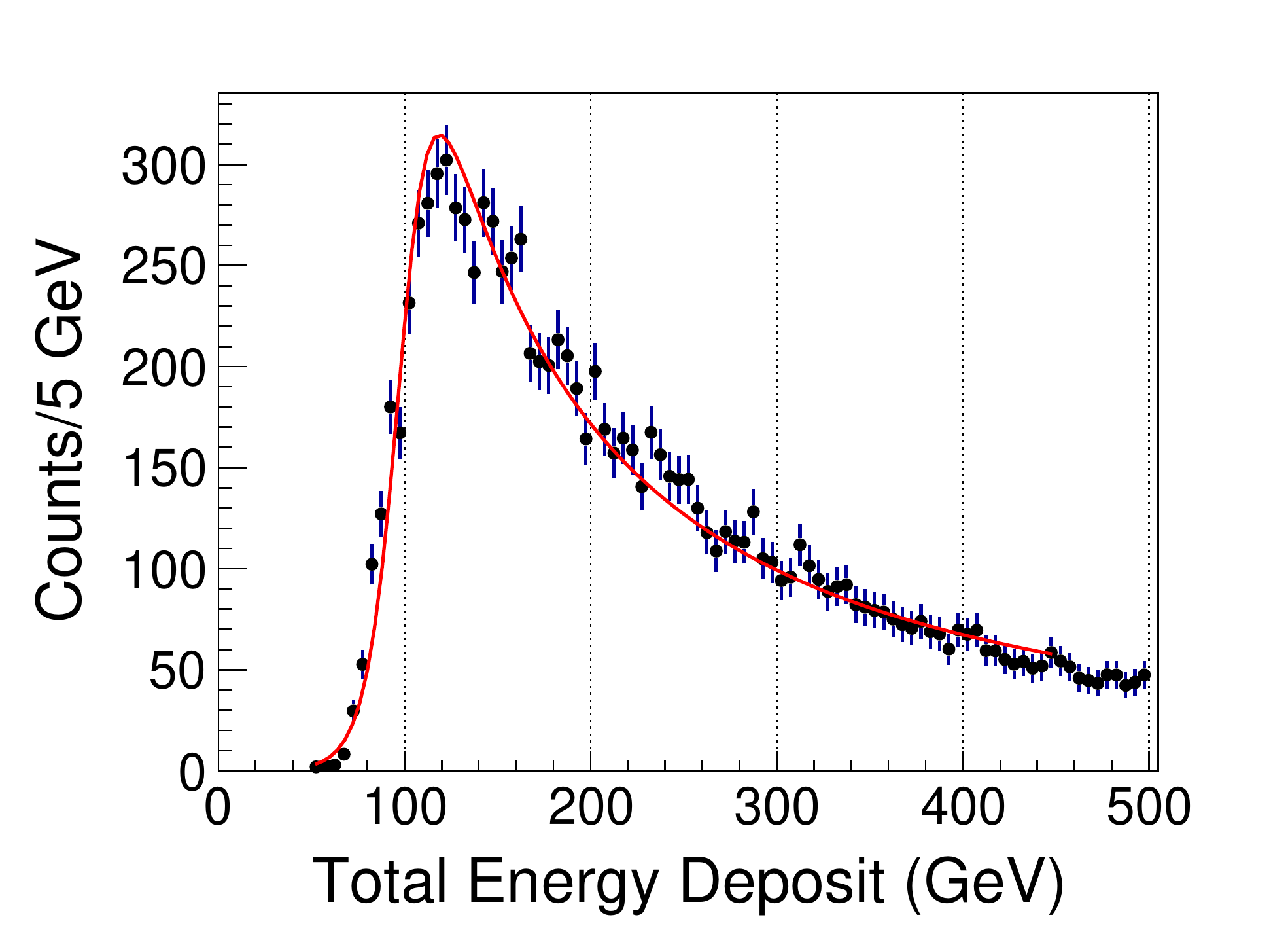}
}

\subfigure[flight data]{
\includegraphics[width=8cm]{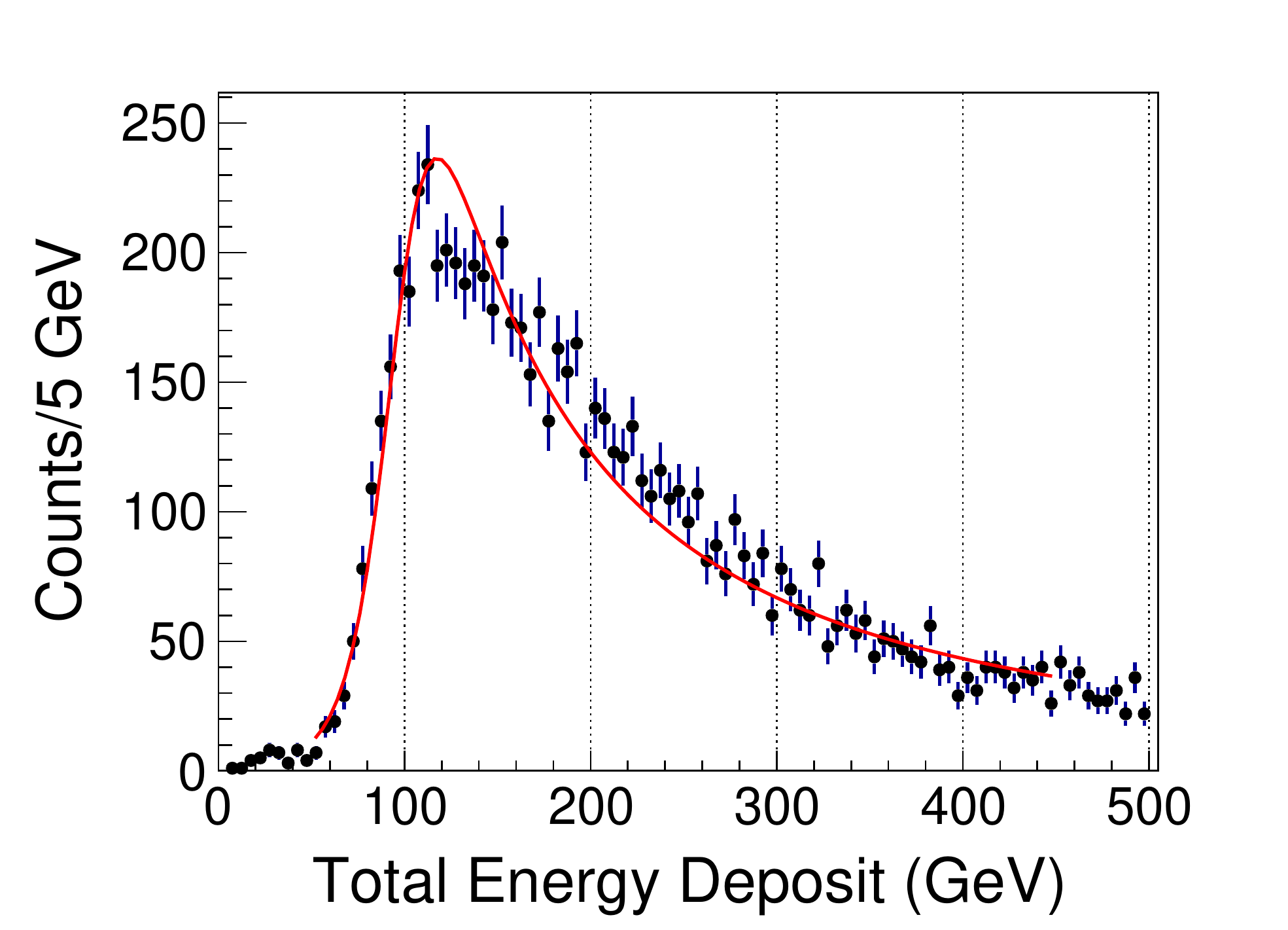}
}
\caption{The total energy deposit by CR iron in the calorimeter. The upper panel is the case for simulation, the lower panel measured data. The dots represent the distribution after the selection procedure. The red line is the fitted function.}
\label{fitting}
\end{figure}

Figure \ref{fitting} shows the fitting on the energy distribution of CR iron for both simulation and flight data. The same procedure was applied to the analyses of the other three elements.

\subsection{Comparison of the fitting parameters}

\begin{figure}[h]
\centering
\includegraphics[width=3.5in]{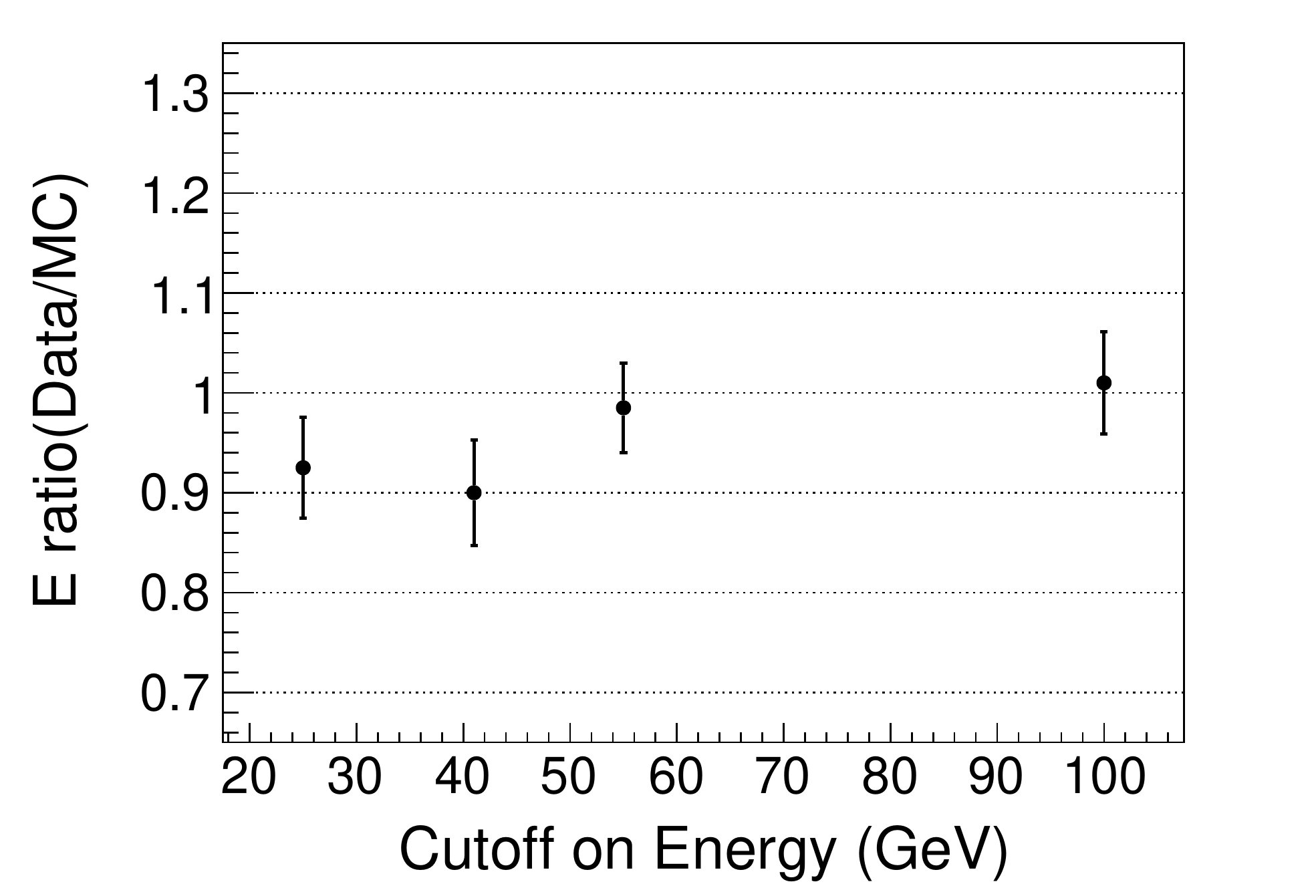}
\caption{The ratio between the energy cutoff on deposited energy and the energy cutoff provided by simulation. Given the uncertainties, the ratios are compatible with one.}
\label{DatavsTracer}
\end{figure}

Since it is the deposited energy spectrum rather than a flux distribution on the kinetic energy, the $\gamma$ parameter does not stand for the so-called "power law" index of flux studies. It was set as a free parameter and differs by $\sim$4\% between the flight data fitting and the simulation fitting for CR iron. The parameter $a$ differs by $\sim$2.5\%. It is even more interesting to compare the $E_c$ parameter. Figure \ref{DatavsTracer} gives the ratio of this parameter in the two spectra.
The systematic uncertainties on this parameter are considered as follows.

For the simulation of back tracing particles in the magnetic field, the satellite experiment HEAO-3 C2 has performed a check on its rigidity cutoff measurement. The HEAO-3 C2 experiment operates during 1979 to 1981 on an orbit of 496 km as altitude and $43.6^{\circ}$ as inclination angle. For its analysis of oxygen nuclei, they found the computed cutoffs by the back tracing method $\sim$3-5 $\pm 2$\% higher than the measured ones. It was deduced that this systematic bias mainly came from the IGRF model \cite{HEAO3C2}\cite{EnergyScaleLAT}. Directly comparing their finding with the work presented here may be difficult, though the two satellites orbit at almost the same altitude. However, the description given by the model is more reliable and accurate near the geomagnetic equator \cite{SMART1994}, which is exactly where the analyses in this paper are performed, as seen in Fig. \ref{mapL}. Therefore it should be safe to assume that the bias caused by the back tracing of particles would be no greater than 3-5\% estimated by HEAO-3 C2 experiment. Their finding could serve as a conservative estimation for the bias of back tracing method while the definitive value of this bias is unknown.

The uncertainty from the Geant4 simulation is estimated according to the nuclei beam test of DAMPE. In the beam test performed in 2015 on H8 beamline at the CERN Super Proton Synchrotron (SPS) facility, the calorimeter response to nuclei of atomic number under 18 were studied \cite{wei2019performance}. The analyses found that the deviation of Geant4 simulation from the beam test data is 1.831\% for carbon, 1.721\% for neon and 0.213\% for silicon at 40 GeV/n. The paper does not give the specific value for iron, since the iron has atomic number higher than 18. However, it is evident that the deviation has a tends to be consistant with only the deviation of helium and lithium larger than the others. Only as an extrapolation, the value for iron might be roughly around 1\%.

For the event selection, the uncertainty resulting from the track reconstruction is estimated as 2.5\% for carbon, 3\% for neon, 2\% for silicon, and 3\% for iron. The uncertainties from other selections such as trigger efficiency are much less and negligible.

\section{Conclusion}

The Earth's magnetic field shields the Earth from cosmic rays of low energy. This fact leads to the geomagnetic cutoff in the energy spectrum of cosmic rays, which can be used as a source to test the response of the BGO calorimeter. By choosing CR carbon, neon, silicon and iron in the data of DAMPE experiment from Jan. 2016 to July 2017, we investigated the response of the BGO calorimeter. Through a data selection, the layer energy and total energy deposits are compared between flight data and simulations. From different perspectives, the energy deposition in the BGO calorimeter exhibits a consistency between simulations and flight data. For the evaluation of the cutoff in the deposited energy spectrum, given the uncertainties, flight data generally are compatible with simulations. Especially for Fe, the two differ only by 1\%, while for C and Ne the difference is larger. All the discrepancies of the four elements in analysis are less than 10\%. 


%



\section*{Acknowledgment}

This work is supported by the National Natural Science Foundation of China (No. 11673021, 11705197), the Joint Funds of the National Natural Science Foundation of China (No, U1738208, U1738139, U1738135) and the National Key Research and Development Program of China (2016YFA0400200).

\ifCLASSOPTIONcaptionsoff
  \newpage
\fi



%

\bibliographystyle{IEEEtran}
\bibliography{manuscript}

%








\end{document}